\documentstyle[epsf]{article}

\def\abstract#1{\vskip 7mm
        \begin{center}{\large Abstract}\par \smallskip
                \begin{minipage}[c]{12cm}
                        \small #1
                \end{minipage}
        \end{center}
}
\def\title#1{\begin{center}{\Large\bf #1}\end{center}}
\def\author#1{\vskip 5mm \begin{center}{#1}\end{center}}
\def\address#1{\begin{center}{\it #1}\end{center}}
\makeatletter
\@ifundefined{lesssim}{}{}
\@ifundefined{gtrsim}{}{}
\def\vereq#1#2{\lower3pt\vbox{\baselineskip1.5pt \lineskip1.5pt
\ialign{$\m@th#1\hfill##\hfil$\crcr#2\crcr\sim\crcr}}}
\makeatother

\begin{document}

\title{%
  Black holes and traversible wormholes: a synthesis
}
\author{%
  Sean A. Hayward\footnote{E-mail: hayward@mm.ewha.ac.kr}
}
\address{%
  Department of Science Education, Ewha Womans University, \\
  Seoul 120-750, Korea
}

\abstract{ A unified framework for black holes and traversible wormholes is
described, where both are locally defined by outer trapping horizons, two-way
traversible for wormholes and one-way traversible for black or white holes. In
a two-dimensional dilaton gravity model, examples are given of: construction of
wormholes from black holes; operation of wormholes for transport, including
back-reaction; maintenance of an operating wormhole; and collapse of wormholes
to black holes. In spherically symmetric Einstein gravity, several exotic
matter models supporting wormhole solutions are proposed: ghost scalar fields,
exotic fluids and pure ghost radiation. }

\section{Introduction}

Space-time wormholes, short cuts between otherwise distant or even unconnected
regions of the universe, are now a familiar plot device in science fiction. As
a theoretical possibility in General Relativity, they gained some scientific
respectability after the article of Morris \& Thorne \cite{MT}. Apart from
attracting public interest and young people to the field, one serious
motivation is that wormholes can increase our understanding of gravity when the
usual energy conditions are not satisfied, due to quantum effects such as the
Casimir effect or Hawking radiation, or in alternative gravitational theories,
such as the recently fashionable brane-world models.

\section{Wormholes and black holes}

\begin{figure}
\centerline{\epsfxsize=15cm \epsfbox{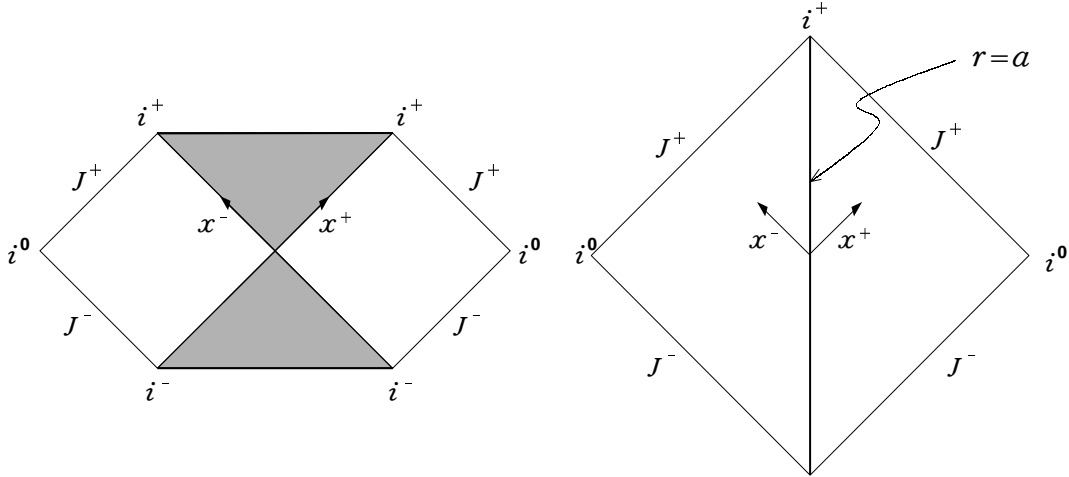}}\vspace{-15cm} \caption{Penrose
diagrams of (i) the Schwarzschild black-hole space-time, (ii) a Morris-Thorne
wormhole space-time. The black-hole horizons and the wormhole throat are outer
trapping horizons, respectively null and time-like. } \label{static}
\end{figure}

The author's interest in wormholes originated with the realization that they
are very similar to black holes, if one thinks of local properties, rather than
the global properties which are usually used to define them. Global
traversibility is incompatible with event horizons, leading to the widespread
view, even among proponents \cite{MT}, that wormholes are quite distinct from
black holes. However, in terms of local properties, both are characterized by
the presence of marginal (marginally trapped) surfaces, and indeed may be
defined by outer trapping horizons, types of hypersurface foliated by marginal
surfaces \cite{bhd,1st,wh}. For a static black hole, the event horizons or
Killing horizons are examples of outer trapping horizons, and for a static
wormhole, the wormhole throat is an example of a double outer trapping horizon,
composed of doubly marginal surfaces (Fig.\ref{static}). The spatial topology
of standard black-hole solutions and Morris-Thorne (static, spherically
symmetric) wormholes is the same, $R\times S^2$, and the spatial geometry can
be identical, as for the Schwarzschild black hole and the spatially
Schwarzschild wormhole \cite{MT}. In each case, a minimal surface connects two
asymptotically flat regions.

The key difference is the causal nature of the trapping horizons. This,
however, is locally determined by the field equations and so may change with
time. Summarizing an earlier proposal \cite{wh}, black holes and wormholes may
be locally defined by outer trapping horizons which are respectively achronal
(space-like or null) and temporal (time-like). This means that they are
respectively one-way and two-way traversible, as desired in each case. The
Einstein equation then shows that they occur respectively under positive and
negative energy density, specifically referring to the null energy condition.
This means that they are supported respectively by normal matter or vacuum, and
what has been dubbed exotic matter \cite{MT}. Given the preponderance of normal
matter in the universe, this means that they respectively occur naturally and
are unlikely to occur naturally, according to present knowledge. However, the
possibility of constructing wormholes seems to be open, given the widespread
appearance of negative energy densities in quantum field theory. Theoretically,
if exotic matter can exist in sufficient concentrations, wormholes are as much
a prediction of General Relativity as black holes.

A consequence of this synthesis is that black holes and wormholes are
interconvertible. The trapping horizons evolve under positive or negative
energy density, changing their causal type to locally characterize black holes
or wormholes. A wormhole could be converted to a black hole if the supporting
exotic matter disperses, or if normal matter is added. Looking at the geometry,
the double trapping horizon constituting a static wormhole throat will
bifurcate under generic perturbations, forming a trapped region. If the two
horizons eventually become null, enclosing a future trapped region, this would
be a black hole (Fig.\ref{dil}(i)). Conversely, a black hole could be converted
to a wormhole by addition of exotic matter, which causes the two black-hole
horizons to become time-like and, if circumstances allow, merge to form a
wormhole throat (Fig.\ref{dil}(ii)). In fact, if one considers a Schwarzschild
black hole evaporating semi-classically by Hawking radiation, the trapping
horizons do indeed become time-like, due to negative-energy radiation absorbed
by the black hole, created by pair production with the escaping radiation. The
evaporating object is a traversible wormhole by any reasonable definition. This
also suggests a wormhole as the endpoint of Hawking evaporation \cite{wh,HPS}.

Finding concrete examples of wormhole and black-hole interconversion requires a
specific exotic matter model, so that there are field equations to determine
the evolution. Four such models are described below, beginning with a simple
model in which analysis is complete, then more realistic models which are under
development.

\section{Two-dimensional dilaton gravity}

The CGHS two-dimensional dilaton gravity model \cite{CGHS} was known to contain
black-hole solutions analogous to Schwarzschild ones, and to share similar
properties such as cosmic censorship \cite{cc}. Generalizing the model to
include a ghost massless Klein-Gordon field, i.e.\ with the gravitational
coupling taking the opposite sign to normal, leads to static wormhole solutions
analogous to Morris-Thorne ones \cite{HKL}. Moreover, the field equations are
explicitly integrable, so it is possible to set initial data corresponding to
dynamical perturbations of black holes or wormholes, then analytically find the
evolved space-time. Four types of processes have been considered in detail
\cite{HKL}, summarized as follows.

\begin{figure}
\centerline{\epsfxsize=15cm \epsfbox{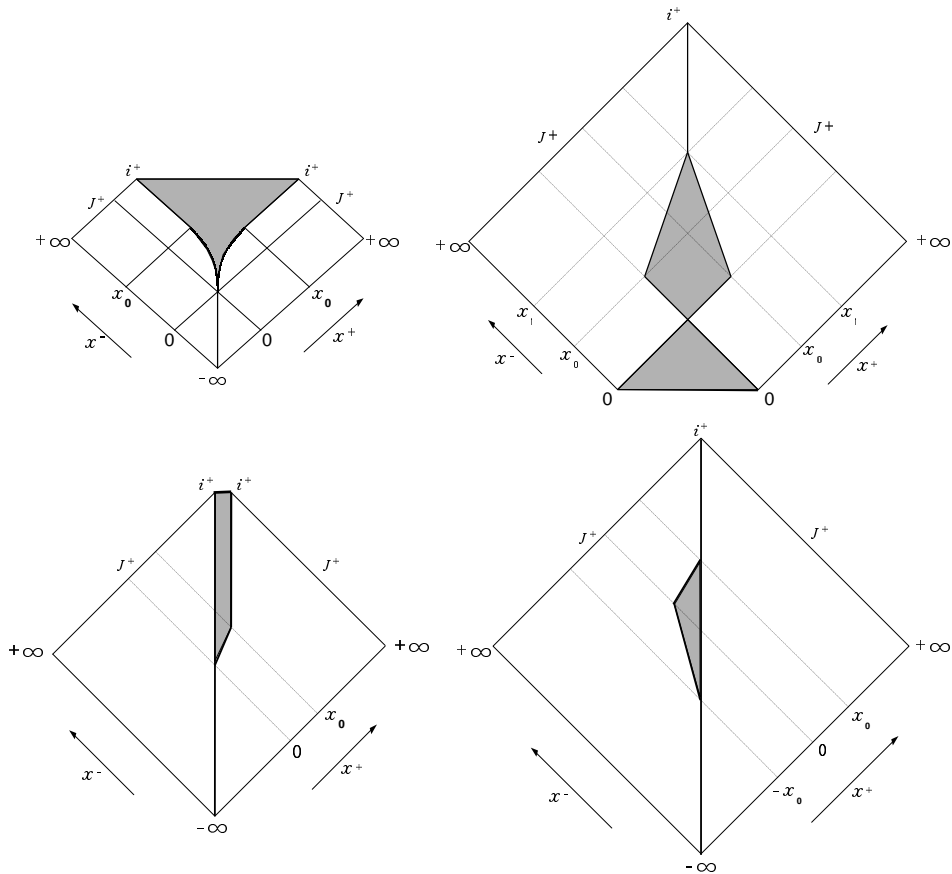}}\vspace{-10cm} \caption{Penrose
diagrams of wormhole (i) collapse to a black hole, (ii) construction from a
black hole, (iii) operation for transport and (iv) maintenance of a static
state, as described in the text. Shading indicates trapped regions. The
trapping horizons evolve under positive or negative energy density, changing
their causal type accordingly. } \label{dil}
\end{figure}

{\bf i. Wormhole collapse} to a black hole. Initial data is set so that there
is initially a static wormhole, with the supporting ghost radiation then
switched off from both sides of the wormhole. The double trapping horizon
bifurcates and each section becomes null, forming a black hole. Solutions were
found for both sudden and gradual collapse (Fig.\ref{dil}(i)).

{\bf ii. Wormhole construction} from a black hole. Initial data is set so that
there is initially a static black hole, which is then irradiated from both
sides with the ghost field. The two null trapping horizons of the black hole
become time-like and, for appropriate radiation profiles, merge to form the
throat of a static wormhole (Fig.\ref{dil}(ii)).

{\bf iii. Wormhole operation} for transport or signalling, including the
back-reaction on the wormhole. Initial data is set so that there is initially a
static wormhole, with a pulse of normal Klein-Gordon radiation then sent
through it. The double trapping horizon again bifurcates, but if the pulse
energy is small, the wormhole remains traversible for a long time
(Fig.\ref{dil}(iii)). This demonstrates the dynamic stability of the static
wormhole.

{\bf iv. Wormhole maintenance} of a static state. As above, but the pulse of
normal radiation is preceded by an extra pulse of ghost radiation with equal
and opposite energy. The double trapping horizon bifurcates and then merges
again, returning the wormhole to its original static state (Fig.\ref{dil}(iv)).

\section{Exotic matter models}

An important shift from the Morris-Thorne approach \cite{MT} is to first
specify an exotic matter model, then discover (suspiciously often) that
wormhole solutions arise naturally. The following exotic matter models are
under study in spherically symmetric Einstein gravity.

{\bf i. Ghost scalar fields}. For the massless ghost Klein-Gordon field, there
is a classic Morris-Thorne wormhole which was actually found previously by
Ellis \cite{E} and several other authors. It has recently been argued to be
stable \cite{AP}. A numerical code for spherically symmetric Einstein gravity
with a massless ghost Klein-Gordon field has been written and used to study
dynamical perturbations of the wormhole \cite{SH}. Preliminary results suggest
that the Ellis wormhole is unstable, with a weakening ghost field causing
collapse to a black hole, and a strengthening ghost field causing an explosion
to an inflating universe.

{\bf ii. Exotic fluids}, i.e.\ fluids not obeying the usual energy conditions.
In particular, the spatially Schwarzschild wormhole \cite{MT} is a solution for
an anisotropic fluid with density zero, radial pressure $-\tau$ and tangential
pressure $\tau/2$, where the tension $\tau$ is positive. This model has
vanishing energy trace, like the Maxwell electromagnetic field. Similar models
have recently been proposed elsewhere \cite{DKMV}.

{\bf iii. Pure ghost radiation}, i.e. pure radiation (or null dust) with
negative energy density. With both ingoing and outgoing radiation, static
wormhole solutions were conjectured in the conference presentation and found
shortly afterwards \cite{pur}.

\section{Conclusion}

A unified theory of black holes and traversible wormholes has been proposed.
Its development includes definitions of mass and surface gravity, and zeroth,
first and second laws of black-hole dynamics and wormhole dynamics
\cite{bhd,1st,wh}. This synthesis is useful even for existing problems such as
black-hole evaporation, where the Hawking radiation converts a Schwarzschild
black hole, at least semi-classically, into a traversible wormhole. It also
raises the largely unexplored issue of wormhole thermodynamics.

Black holes are now generally accepted as astrophysical realities, whereas
traversible wormholes are often regarded as unphysical theoretical curiosities,
outside mainstream scientific research, just as black holes once were. The dual
nature of black holes and wormholes, in particular their dynamic
interconvertibility, forces new viewpoints. Wormholes are just black holes
under negative energy density.

\bigskip\noindent Research supported by Korea Research Foundation grant
KRF-2001-015-DP0095. Thanks to the conference organizers for support and
Hyunjoo Lee for preparing the figures.

\end{document}